\documentclass{elsart}
\usepackage{natbib}
\usepackage{amsmath,amssymb,graphicx}
\begin{document}
\runauthor{Croft, Rawlings, Hill}
\begin{frontmatter}
\title{High-redshift clusters from NVSS:\\The TexOx Cluster (TOC) Survey}
\author[oxford,llnl]{Steve Croft}
\author[oxford]{Steve Rawlings}
\author[texas]{Gary J.~Hill}

\address[oxford]{Astrophysics, University of Oxford, Keble Road, Oxford, OX1 3RH, UK}
\address[texas]{McDonald Observatory, Dept. of Astronomy, University of Texas at Austin, RLM 15.308, TX 78712, USA}
\address[llnl]{IGPP, Lawrence Livermore National Laboratory, L-413, 7000 East Ave., Livermore, CA 94550, USA}
\begin{abstract}
The TexOx Cluster (TOC) Survey uses overdensities of radiosources in the NVSS to trace clusters of galaxies. The links between radiosources and rich environments make this a powerful way to find clusters which may potentially be overlooked by other selection techniques. By including constraints from optical surveys, TOC is an extremely efficient way to find clusters at high redshift. One such field, TOC J0233.3+3021, contains at least one galaxy cluster (at $z \sim 1.4$) and has been detected using the Sunyaev-Zel'dovich (SZ) effect.  Even in targetted deep optical observations, however, distinguishing the cluster galaxies from the background is difficult, especially given the tendency of TOC to select fields containing multiple structures at different redshifts.
\end{abstract}
\end{frontmatter}

\section{Introduction}
Galaxy clusters have traditionally been used as cosmological probes \cite{bf98}, and despite the much heralded recent dawning of the age of precision cosmology, they will remain useful as such in determining the normalisation of the matter power spectrum, the cosmic equation of state \cite{haiman01}, and refining measurements of other cosmological parameters. The current concordance of cosmological constraints, however, now makes it feasible to rely on the redundancy between measurement methods \cite{tz02} to use the predictions of the standard $\Lambda$CDM model to test cluster and AGN physics rather than the reverse.

Clusters, especially at high redshift, are dynamic environments, rife with mergers \cite{bh96}, galaxy harassment \cite{m98}, enhanced star formation \cite{z96}, tidal stripping \cite{go72} and other processes which transform morphology and stellar populations. Many of these processes hinder the identification of clusters, as all methods make implicit assumptions about what a cluster should look like, and then try to find objects fulfilling such criteria, making each method subject to (often ill-understood) selection effects.

Whilst radio selection of clusters does not offer a ``clean'' method by which to construct an unbiased cluster sample, the selection effects are at least different, and the lack of dust obscuration in the radio enables one to probe high redshifts more easily \cite{v02}. Radiogalaxies tend to be located in clusters \cite{hl91,best00} and links between cluster formation and radiogalaxy triggering \cite{sr02} are expected in the dynamic environments noted above. This makes the comparison of radio-selected clusters with those detected by other methods an attractive means by which AGN triggering, fuelling, evolution, duty cycle, obscuration and environment may be studied.

\section{The TexOx Cluster (TOC) Survey}
The cosmic evolution of the radiosource population \cite{w01} suggests that some high-redshift clusters will contain several radio-loud AGN. By targetting imaging observations towards fields containing overdensities of radiosources, we ought to be able to detect such clusters. The TexOx (Texas--Oxford) Cluster Survey utilises this technique by searching for $\sim 7 \sigma$ over-densities at $\sim$ mJy levels in $7^{\prime} \times 7^{\prime}$ boxes within the NVSS radio catalogue \cite{NVSS}, finding 331 within an $1128~\rm deg^2$ sky area \cite{TOC}.

The TOC algorithm allows each source to be a member of no more than one overdensity. This prevents large overdensities from being detected several times, artificially increasing the number of ``real'' clusters detected. Running the TOC algorithm multiple times on random simulations of the same sky area (with the same source density) as one of the regions of the real TOC survey resulted in a Gaussian distribution of number of clusters detected. The mean of this Gaussian, 65.5, was very significantly ($\sim 6 \sigma$) less than the number of clusters, 103, found in the same region of the NVSS.

We conclude, therefore, that at most  $\sim 60 \pm 7\%$ of TOC clusters are chance superpositions of unrelated radiosources. However, many of these fields are expected to contain at least one real cluster due to the high probability that even single radiosources may be found in clusters (and even if all sources are at different redshifts, TOC is $\sim 5$ times more efficient than looking around isolated sources). Inspection of the NVSS radio maps suggests that $\sim 20 \%$ of TOC fields are extended structures resolved by NVSS into several components, leaving $\sim 20 \%$ of fields to contain at least two (and potentially many more) physically-related radiosources. This means that we would expect TOC to discover $\gtrsim 65$ fields containing two or more radiosources at the same redshift.

By targetting those fields where none of the NVSS sources are identified on the POSS-II \cite{POSS} images, we are able to preferentially target high-redshift clusters. Around $30\%$ of NVSS sources have POSS-II counterparts \cite{s02}, so assuming we are able to remove the extended structures mentioned above from the sample, there is only a $17\%$ chance that a given field of 5 radiosources would have no POSS-II identifications if the radiosources were randomly distributed in redshift. This POSS-II constraint (POSS-II can see $L_\star$ galaxies at $z \lesssim 0.3$), the $z \sim 1$ median redshift of NVSS sources, and the fact that the TOC  $7^{\prime} \times 7^{\prime}$ box size is comparable to the size of a cluster core at $z \gtrsim 0.2$, means that TOC is highly efficient at finding high-redshift clusters.

Deep optical imaging (30 -- 60 minutes per band in $R$ and $I$ at the McDonald 2.7-m in around 130 fields) was used to select around 20 promising high-redshift candidates for near-IR followup in $J$ and $H$ at the Calar Alto 3.5-m. Additionally some $K$ data was obtained at UKIRT, and deeper, higher-resolution radio maps with the VLA.

\section{TOC~J0233.3+3021 - a high redshift cluster?}

One particularly promising high-redshift cluster candidate was \linebreak TOC~J0233.3+3021, a field in which the radiosources were identified with galaxies of $K \approx 18$, $J \approx 20$, $R \approx 24$. Our collaborators in Cambridge observed this field with the Ryle Telescope and measured a 15~GHz SZ decrement \cite{sz} of $-695 \mu$Jy, at $7\sigma$ significance, consistent with the presence of a cluster with a minimum gas mass of $5 \times 10^{13} M_{\odot}$ \cite{cotter02}.

The magnitudes of the radiogalaxies strongly suggest the presence of a cluster at $z \gtrsim 1$; attempts to obtain spectroscopic redshifts of the NVSS identifications using ISIS on the WHT resulted in the detection of faint red continuum in the four brightest cases, but no evidence of emission lines. We used the publicly available software {\sc hyperz} \cite{hyperz} to obtain photometric redshifts using our $BRIJHK$ photometry and determined best-fit redshifts of $z \sim 1.4$ in four cases, and $z \sim 1.0$ for the remaining radiogalaxy. The best solutions were $3 - 6 L_\star$ elliptical templates in the former cases, and a $2 L_\star$ S0 in the latter. One of the ellipticals has a poorly constrained redshift and is consistent with being at $z \sim 1.0$.

\begin{figure}
\centering
\includegraphics[width=2.7in]{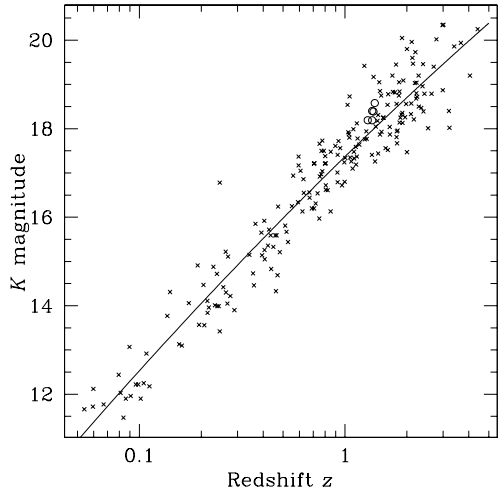}
\caption{\label{fig:kz}The $K - Z$ relation, showing the data for 7C-I, 7C-II,  7C-III, 6CE, 6C$^{\star}$ and 3CRR (crosses) and the best-fit between $K$ and $\log_{10} z$ from Willott et al.\,\cite{willott:kz}. Open circles are the radiogalaxies from TOC~J0233.3+3021, using photometric redshifts.}
\end{figure}

The consistency of the photometric redshifts suggests that at least three of the radiogalaxies may be members of a structure at $z \sim 1.4$; this is supported by their position on the radiogalaxy $K-z$ diagram (Fig.~\ref{fig:kz}). From the integrated 1.4~GHz radio fluxes, we obtain rest-frame 1.4~GHz luminosities\footnote{Assuming a spectral index $\alpha = 0.8$, where $L_\nu \propto \nu^{-\alpha}$} of $\sim 10^{25}$~W~Hz$^{-1}$~sr$^{-1}$, close to the FR-I/II break \cite{lo96}. One of the ellipticals hosts a radiosource which is clearly an FR-II from our 8~GHz VLA maps (with a projected size of 140~kpc); the rest are compact doubles or are unresolved. Two of the sources appear to have distorted FR-I type morphologies, as might be expected if they were in clusters, or perhaps groups falling into a cluster.

\begin{figure}
  \begin{minipage}[t]{2.7in}
    \includegraphics[width=2.5in]{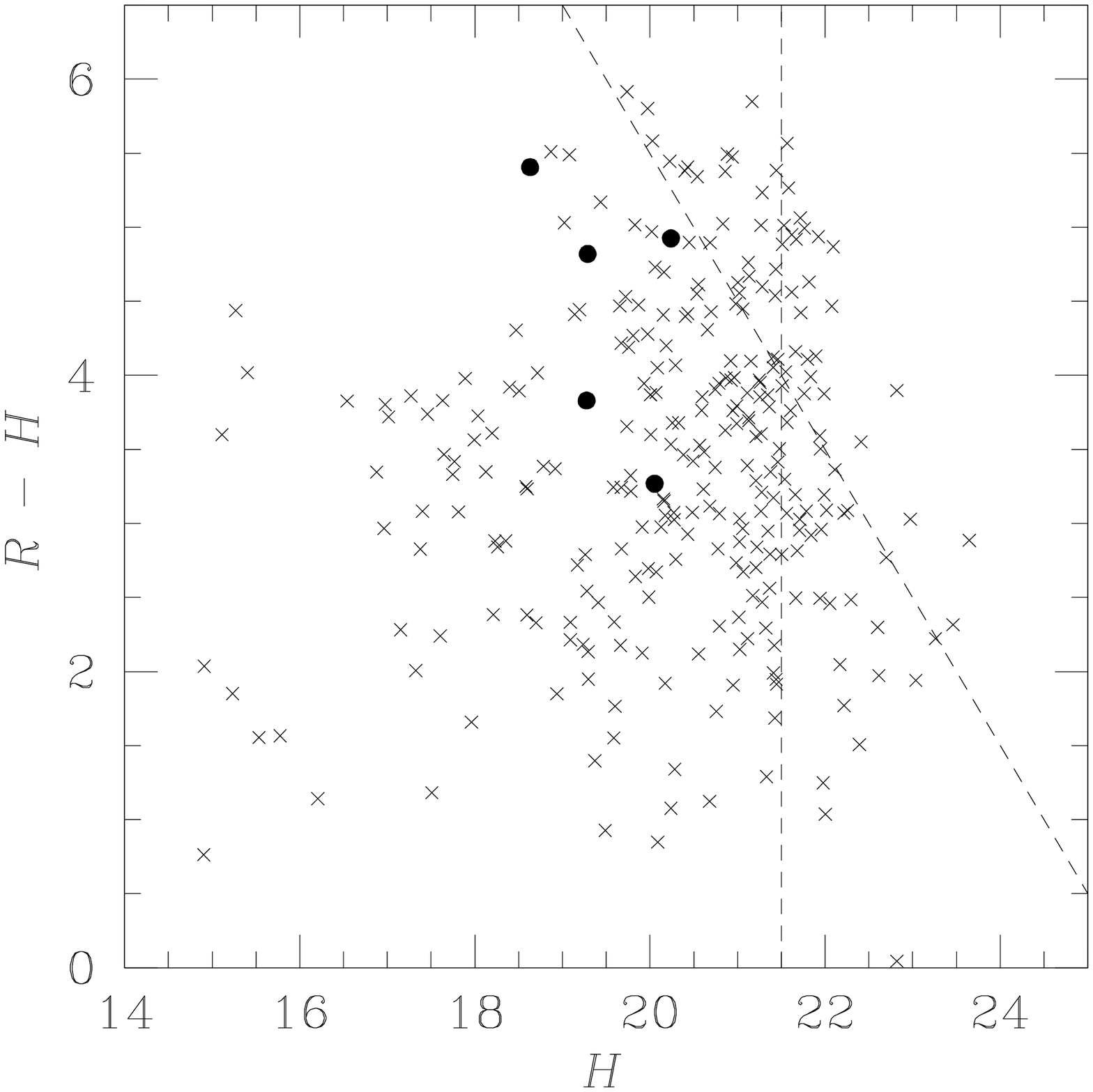}
\parbox{0.95\textwidth}{\caption{\label{fig:cm}Colour-magnitude diagram for $R$ and $H$ detections in the field of TOC~J0233.3+3021 (crosses). The NVSS radiogalaxies in the field are the filled circles. The dashed lines represent completeness limit estimates ($H \sim 21.5$ and $R \sim 24.5$).}}
  \end{minipage}
  \begin{minipage}[t]{2.7in}
    \includegraphics[width=2.5in]{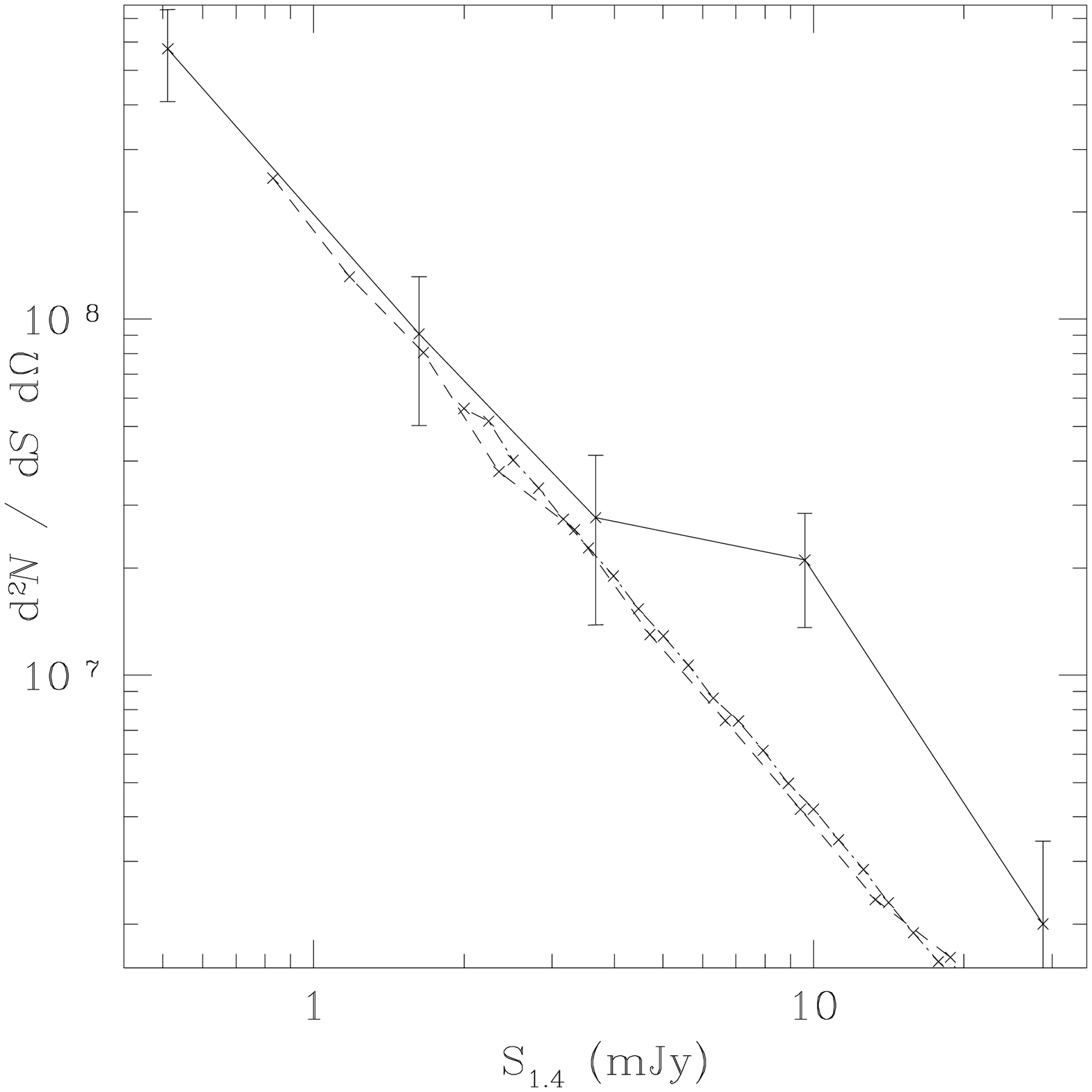}
\parbox{0.95\textwidth}{\caption{\label{fig:rsrc} Differential 1.4~GHz radio source counts in the field of TOC~J0233.3+3021 (sources Jy$^{-1}$ sr$^{-1}$), compared with those from FIRST \cite{first} and ATESP \cite{atesp} -- dot-dashed and dashed lines respectively.
}}
  \end{minipage}
\end{figure}

Although it seems probable that the SZ signal is associated with the inferred overdensity of radiogalaxies at $z \sim 1.4$, it is of course possible that it is associated with a cluster at a different redshift. In an attempt to detect the signature of the presence of a cluster we are substantially aided by the availability of deep radio, optical and IR data, but hampered by the lack of spectroscopic redshifts. In an attempt to identify a cluster red sequence, we plotted a colour-magnitude diagram for those objects in the field with both $R$ and $H$ detections and magnitude errors $< 0.3$. The resulting 297 objects are shown in Fig.~\ref{fig:cm}. There is perhaps a hint of two red sequences; the redder contains 3 of the radiosources and is similar in colour and magnitude to that seen by Rosati et al.\,\cite{rosati99} in a $z = 1.26$ cluster. The lower redshift red sequence contains the other two radiosources (consistent with the photometric redshift analysis).

Number counts of objects from our Gemini GMOS image of this field show a slight excess over galaxy number counts obtained by other authors \cite{postman98} at $i^\prime \sim 20$, possibly associated with the inferred $z \sim 1.0$ structure, but the presence of a cluster at $z \sim 1.4$ is almost impossible to ascertain from these data due to the faintness of the objects, the large surface densities of background galaxies, and a drop in completeness in our data at these magnitudes. Radio number counts, however, show an excess of sources, both at the several mJy level at which the overdensity was selected (hardly surprising!) and, very tentatively, at fainter flux densities, before the contrast becomes lost in the rising background source counts (Fig~\ref{fig:rsrc}).

\section{Summary}

The TOC method provides an efficient way to find clusters over a wide range of redshift, out to $z > 1$. The contrast provided against the background radiosource counts at mJy levels makes the selection relatively easy, but distinguishing the cluster population from background and foreground sources will require a good deal of effort, ideally using multi-object spectroscopy. We note that the difficulties of clearly identifying clusters at high-redshift, especially in the ``redshift desert'' ($1.2 \lesssim z \lesssim 2$) compounds the problem; indeed other authors \cite{haynes02} have also had difficulties identifying the systems responsible for SZ decrements in some fields.

 Our technique suffers from the problem that it tends to find superpositions of systems at different redshifts, complicating matters further. This may prove to be an advantage, however, should one wish to study the population of a more distant cluster using a foreground cluster as a gravitational lens.

\paragraph*{Acknowledgments}

We thank N.\ Drory, our Calar Alto collaborator; G.\ Cotter, H.\ Buttery, R.\ Das, K.\ Grainge, W.\ Grainger, M.\ Jones, G.\ Pooley, and R.\ Saunders, our Cambridge SZ collaborators; and J.\ Tufts and P.\ Gay for help with observing. The work of SC was performed with the support of a PPARC PhD studentship, and under the auspices of the U.S. Department of Energy, National Nuclear Security Administration by the University of California, Lawrence Livermore National Laboratory under contract No. W-7405-Eng-48. This material is based in part upon work supported by the Texas Advanced Research Program under Grant No. 009658-0710-1999.

\end{document}